\documentclass[]{svmult}
\usepackage{makeidx}
\usepackage{multicol}
\newif\ifpretentious \pretentioustrue	
\usepackage{amsfonts}			
\usepackage[OT2,OT1]{fontenc}		
\ifpretentious
  \newcommand\cyr{%
  \renewcommand\rmdefault{wncyr}
  \renewcommand\bfdefault{b}
  \renewcommand\encodingdefault{OT2}%
  \normalfont\selectfont}%
  \DeclareTextFontCommand{\textcyr}{\cyr}
  \DeclareTextSymbol{\cprime}{OT2}{'176} 
\else
  \def\ss{ss}				
  \def\textcyr{}			
  \def\cprime{'}			
\fi
\newif\ifpdf
\ifx\pdfoutput\undefined\pdffalse\else\pdfoutput=1\pdftrue\fi
\ifpdf
  \usepackage[backref,citecolor=blue]{hyperref}
  \let\myhref=\href\def\href#1#2{\penalty-20\myhref{#1}{\tt #2}}%
\else%
  \def\href#1#2{{\penalty-20\tt #2}}
\fi

\usepackage{epsfig}
\def\rational#1#2{{\mathchoice{\textstyle{#1\over#2}}%
  {\scriptstyle{#1\over#2}}{\scriptscriptstyle{#1\over#2}}{#1/#2}}}
\def\half{\rational12}			
\def\third{\rational13}			
\def\quarter{\rational14}		
\def\C{{\mathbb C}}			
\def\R{{\mathbb R}}			
\def\N{{\mathbb N}}			
\def\Z{{\mathbb Z}}			
\def\defn{\equiv}			
\def\prd{\omega}			
\def\res{\mathop{\rm Res}}		
\def\wq{Q\index{Weierstrass Q function}} 
\let\wpchar=\wp
\def\wp{\relax\wpchar\index{Weierstrass elliptic function}}
\def\wzeta{\zeta\index{Weierstrass zeta function}}
\def\wsigma{\sigma\index{Weierstrass sigma function}}
\def\sn{\mathop{\rm sn}\index{Jacobi elliptic function sn}} 
\def\cn{\mathop{\rm cn}\index{Jacobi elliptic function cn}} 
\def\dn{\mathop{\rm dn}\index{Jacobi elliptic function dn}}
\def\K{K\index{Complete elliptic integral}} 
\def\sgn{\mathop{\rm sgn}\index{Signum function}} 
\def\union{\cup}			
\def\implies{\Rightarrow}		
\def\refsec#1{(\S\ref{#1})}		
\def\zolotarev{\textcyr{Zolotarev}\index{Zolotarev}}
\def\chebyshev{\textcyr{Chebyshev}\index{Chebyshev}}
\def\gauss{{Gau\ss}\index{Gauss}}
\def\weierstrass{{Weierstra\ss}\index{Weierstrass}}
\def\liouville{Liouville\index{Liouville}}
\def\cauchy{Cauchy\index{Cauchy}}
\def\jacobi{Jacobi\index{Jacobi}}

\makeindex

\begin{document}

\title*{Fast Evaluation of {\ifpdf Zolotarev\else\zolotarev\fi} Coefficients}
\author{A. D. Kennedy}
\institute{School of Physics, University of Edinburgh, EH9 3JZ, Scotland, UK 
  \texttt{adk@ph.ed.ac.uk}}
\maketitle

\abstract{\noindent  We review  the  theory of  elliptic  functions leading  to
\zolotarev's formula for the sign function over the range \(\varepsilon \leq|x|
\leq1\). We show  how \gauss' arithmetico-geometric mean allows  us to evaluate
elliptic functions cheaply, and  thus to compute {\zolotarev} coefficients ``on
the fly'' as a function of \(\varepsilon\). This in turn allows us to calculate
the matrix  functions \(\sgn H\), \(\sqrt  H\), and \(1/\sqrt  H\) both quickly
and  accurately for  any  Hermitian matrix  \(H\)  whose spectrum  lies in  the
specified range.\parfillskip=0pt\par}

\section{Introduction}

The purpose of  this paper is to  provide a detailed account of  how to compute
the coefficients of \zolotarev's optimal rational approximation to the \(\sgn\)
function. This is  of considerable interest for lattice  QCD because evaluation
of  the  Neuberger  overlap operator  \cite{Neuberger:1997fp,Neuberger:1998my,%
Edwards:1998yw,Borici:2002a}  requires  computation  of the  \(\sgn\)  function
applied  to  a Hermitian  matrix~\(H\).  Numerical  techniques  for applying  a
rational  approximation  to  a  matrix  are  discussed  in  a  companion  paper
\cite{kennedy:2003a}, and in \cite{rivlin:1981,achieser:1956}.

In general, the computation of optimal (\chebyshev) rational approximations for
a continuous function  over a compact interval requires  an iterative numerical
algorithm \cite{remez:1957,cheney:1966},  but for the function  \(\sgn H\) (and
the related  functions \(\sqrt H\) and  \(1/\sqrt H\)~\cite{kennedy:2003a}) the
coefficients   of    the   optimal   approximation   are    known   in   closed
form in terms of Jacobi elliptic functions~\cite{zolotarev:1877}.

We   give   a   fairly   detailed   summary   of   the   theory   of   elliptic
functions~\refsec{sec:elliptic}  \cite{achieser:1990,whittaker:1927} leading to
the  principal modular  transformation  of degree~\(n\)~\refsec{sec:principal},
which  directly  leads   to  \zolotarev's  formula~\refsec{sec:zolotarev}.  Our
approach closely follows that presented in~\cite{achieser:1990}.

We also explain how to evaluate the elliptic functions necessary to compute the
{\zolotarev}  coefficients~\refsec{sec:evaluate},  explaining  the use  of  the
appropriate   modular  transformations~\refsec{sec:modular}   and   of  \gauss'
arithmetico-geometric mean~\refsec{sec:agmean},  as well as  providing explicit
samples of code for the latter~\refsec{sec:implement}.

\section{Elliptic Functions} \label{sec:elliptic}

\subsection{Introduction}

There   are   two   commonly   encountered   types   of   elliptic   functions:
\weierstrass~\refsec{sec:weierstrass}      and      \jacobi~\refsec{sec:jacobi}
functions. In  principle these  are completely equivalent:  indeed each  may be
expressed   in  terms   of  the   other~\refsec{sec:repsn};  but   in  practice
{\weierstrass}  functions  are  more  elegant  and natural  for  a  theoretical
discussion, and {\jacobi}  functions are the more convenient  for numerical use
in most applications.

Elliptic     functions     are     doubly     periodic     complex     analytic
functions~\refsec{sec:periodic};  the  combination  of  their  periodicity  and
analyticity  leads to  very strong  constraints on  their structure,  and these
constraints   are    most   easily    extracted   by   use    of   \liouville's
theorem~\refsec{sec:liouville}. The constraints imply  that for a fixed pair of
periods \(\prd\) and \(\prd'\) an  elliptic function is uniquely determined, up
to an  overall constant  factor, by the  locations of  its poles and  zeros. In
particular,  this  means  that  any   elliptic  function  may  be  expanded  in
\emph{rational  function}  or \emph{partial  fraction}  form  in  terms of  the
{\weierstrass}~\refsec{sec:expansion}  function with the  same periods  and its
derivative. Furthermore, this in turn shows that all elliptic functions satisfy
an \emph{addition theorem}~\refsec{sec:addition} which allows us to write these
expansions    in   terms   of    {\weierstrass}   functions    with   unshifted
argument~\(z\)~\refsec{sec:representation}.

There  are many  different choices  of  periods that  lead to  the same  period
lattice, and this representation theorem allows  us to express them in terms of
each other:  such transformations  are called \emph{modular  transformations of
degree one}. We may also specify periods whose period lattice properly contains
the original period  lattice; and elliptic functions with  these periods may be
represented rationally  in terms of  the original ones.  These (non-invertible)
transformations are  \emph{modular transformations}  of higher degree,  and the
set of all modular transformations form  a semigroup that is generated by a few
basic transformations  (the {\jacobi}  real and imaginary  transformations, and
the principal transformation of degree~\(n\),  for instance). The form of these
modular  transformations  may be  found  using  the  representation theorem  by
matching the location  of the poles and zeros of the  functions, and fixing the
overall constant at some suitable position.

One  of  the periods  of  the {\weierstrass}  functions  may  be eliminated  by
rescaling the argument,  and if we accept this  trivial transformation then all
elliptic   functions   may   be   expressed   in   terms   of   the   {\jacobi}
functions~\refsec{sec:jacobi} with a single parameter~\(k\).

In order to evaluate the  {\jacobi} functions for arbitrary argument and (real)
parameter we may  first use the {\jacobi} real transformation  to write them in
terms of {\jacobi} functions whose parameter lies in the unit interval; then we
may use the addition theorem to write them in terms of functions of purely real
and imaginary arguments, and finally use the {\jacobi} imaginary transformation
to rewrite the  latter in terms of functions with real  arguments. This can all
be done numerically or analytically, and is explained in detail for the case of
interest in~\refsec{sec:numeval}.

We  are left  with  the problem  of  evaluating {\jacobi}  functions with  real
argument and parameter in the unit  interval. This may be done very efficiently
by use of \gauss' method of the arithmetico-geometric mean~\refsec{sec:agmean}.
This makes use  of the particular case of  the principal modular transformation
of  degree~2, known  as the  \emph{{\gauss}  transformation} to  show that  the
mapping \((a,b) \mapsto \left(\half(a+b),\sqrt{ab}\right)\) iterates to a fixed
point; for a suitable choice of the initial values the value of the fixed point
gives us  the value of the  complete elliptic integral~\(\K\), and  with just a
little more effort it can be induced to give us the values of all the {\jacobi}
functions too~\refsec{sec:numeval}.

This  procedure  is sufficiently  fast  and accurate  that  the  time taken  to
evaluate the coefficients of  the {\zolotarev} approximation for any reasonable
values  of the  range  specified  by \(\varepsilon\)  and  the degree~\(n\)  is
negligible compared to the cost of applying the approximate operator \(\sgn H\)
to a vector.

\subsection{Periodic functions} \label{sec:periodic}

A function \(f:\C\to\C\) is \emph{periodic} with \emph{period} \(\prd\in\C\) if
\(f(z) = f(z  + \prd)\). Clearly, if \(\prd_1, \prd_2,  \ldots\) are periods of
\(f\) then any  linear combination of them with  integer coefficients, \(\sum_i
n_i\prd_i\), is also a period; thus the periods form a \(\Z\)-module.

It is obvious that  if \(f\) is a constant then this  \(\Z\)-module is dense in
\(\C\),  but the converse  holds too,  for if  there is  a sequence  of periods
\(\prd_1, \prd_2, \ldots\)  that converges to zero, then  \(f'(z) = \lim_{n \to
\infty}  \left[f(z+\prd_n) - f(z)\right]/\prd_n  = 0\).  It follows  that every
non-constant function must have a set of \emph{primitive periods}, that is ones
that  are  not sums  of  integer multiples  of  periods  of smaller  magnitude.
{\jacobi}  showed that  if  \(f\)  is not  constant  it can  have  at most  two
primitive periods, and that these two periods cannot be colinear.

\subsection{\liouville's Theorem} \label{sec:liouville}

From  here on  we shall  consider only  doubly periodic  meromorphic functions,
which  for  historical  reasons  are called  \emph{elliptic  functions},  whose
non-colinear primitive  periods we shall call \(\prd\)  and \(\prd'\). Consider
the integral of  such a function \(f\) around  the parallelogram \(\partial P\)
defined by its primitive periods,  \[\oint_{\partial P} dz\, f(z) = \int_0^\prd
dz\, f(z)  + \int_\prd^{\prd+\prd'} dz\, f(z)  + \int_{\prd+\prd'}^{\prd'} dz\,
f(z) + \int_{\prd'}^0 dz\, f(z).\] Substituting \(z' = z - \prd\) in the second
integral and \(z'' = z - \prd'\) in the third we have \[\oint_{\partial P} dz\,
f(z) = \int_0^\prd dz\, \left[f(z)  - f(z + \prd')\right] + \int_0^{\prd'} dz\,
\left[f(z  + \prd)  - f(z)\right]  = 0,\]  upon observing  that  the integrands
identically vanish  due to the periodicity  of \(f\). On the  other hand, since
\(f\) is meromorphic we can evaluate it  in terms of its residues, and hence we
find that the sum  of the residues at all the poles of  \(f\) in \(P\) is zero.
Since the sum of the residues at all the poles of an elliptic function are zero
an elliptic function cannot have  less than two poles, taking multiplicity into
account.

Several useful  corollaries follow immediately from this  theorem. Consider the
logarithmic derivative \(g(z)  = [\ln f(z)]' = f(z)'/f(z)\)  where \(f\) is any
elliptic function which is not  identically zero. We see immediately that \(g\)
is  holomorphic everywhere  except at  the discrete  set  \(\{\zeta_j\}\) where
\(f\) has  a pole  or a zero.  Near these  singularities \(f\) has  the Laurent
expansion   \(f(z)   =  c_j(z-\zeta_j)^{r_j}   +   O\left(  (z-   \zeta_j)^{r_j
+1}\right)\) with  \(c_j\in\C\) and  \(r_j\in\Z\), so the  residue of  \(g\) at
\(\zeta_j\)  is \(r_j\).  Applying the  previous result  to the  function \(g\)
instead of  \(f\) we find that \(\oint_{\partial  P} dz\, g(z) =  2\pi i \sum_j
r_j = 0\), or  in other words that the number of poles  of \(f\) must equal the
number of zeros of \(f\), counting multiplicity in both cases.

It  follows immediately  that there  are no  non-constant  holomorphic elliptic
functions; for if  there was an analytic elliptic function  \(f\) with no poles
then \(f(z) - a\) could have no zeros either.

If we consider the function \(h(z) = zg(z)\) then we find 
\begin{eqnarray*}
\oint_{\partial P} dz\, h(z) &=& \int_0^\prd dz\, h(z) + \int_\prd^{\prd+\prd'}
dz\, h(z) +  \int_{\prd+\prd'}^{\prd'} dz\, h(z) + \int_{\prd'}^0  dz\, h(z) \\
&=&  \int_0^\prd  dz\, \left[h(z)  -  h(z+\prd')\right]  + \int_0^{\prd'}  dz\,
\left[h(z+\prd) -  h(z)\right] \\ &=& \int_0^\prd dz\,  \left[zg(z) - (z+\prd')
g(z)\right] +  \int_0^{\prd'} dz\, \left[(z+\prd)g(z)  - zg(z)\right] \\  &=& -
\prd' \int_0^\prd  dz\, g(z)  + \prd  \int_0^{\prd'} dz\, g(z)  \\ &=&  - \prd'
\left\{\ln[f(\prd) - a]  - \ln[f(0) - a]\right\} +  \prd \left\{ \ln[f(\prd') -
a] - \ln[f(0) - a]\right\} \\ &=& 2\pi i(n'\prd' + n\prd),
\end{eqnarray*}
where \(n,n'\in\N\) are the number of times \(f(z)\) winds around the origin as
\(z\) is taken along the straight  line from \(0\) to \(\prd\) or \(\prd'\). On
the other hand, \cauchy's theorem  tells us that \[\oint_{\partial P} dz\, h(z)
= \oint_{\partial S} {dz\, zf'(z)\over  f(z)} = 2\pi i \sum_{k=1}^m (\alpha_k -
\beta_k),\] where \(\alpha_k\)  and \(\beta_k\) are the locations  of the poles
and zeros  respectively of \(f(z)\), again  counting multiplicity. Consequently
we have that \(\sum_{k=1}^m (\alpha_k -  \beta_k) = n\prd + n'\prd'\), that is,
the sum  of the locations  of the poles  minus the sum  of the location  of the
zeros of any elliptic function is zero modulo its periods.

\subsection{{\weierstrass} elliptic functions} \label{sec:weierstrass}

The most  elegant formalism  for elliptic functions  is due to  \weierstrass. A
simple  way  to construct  a  doubly periodic  function  out  of some  analytic
function \(f\) is to construct the  double sum \(\sum_{m,m'\in\Z} f(z - m\prd -
m'\prd)\).  In  order for  this  sum to  converge  uniformly  it suffices  that
\(|f(z)| < k/z^3\), so a simple choice is \(\wq(z) \defn -2 \sum_{m,m'\in\Z} (z
- m\prd - m'\prd')^{-3}\). Clearly this  function is doubly periodic, \(\wq(z +
\prd) = \wq(z + \prd') = \wq(z)\), and odd, \(\wq(-z) = -\wq(z)\).

The derivative  of an elliptic function  is clearly also  an elliptic function,
but  in  general the  integral  of  an elliptic  function  is  not an  elliptic
function. Indeed, if we define the {\weierstrass} \(\wp\) function\footnote{The
name  of the  function is  \(\wp\), but  I do  not know  what the  name  of the
function is  called; q.v.,  ``Through the Looking-Glass,  and what  Alice found
there,'' Chapter VIII, p.~306, footnote~8 \cite{caroll:1993}.} such that \(\wp'
= \wq\) we know that \(\wp(z + \prd) = \wp(z) + c\) for any period \(\prd\). In
this  case we also  know that  \(\wp\) must  be an  even function,  \(\wp(-z) =
\wp(z)\), because \(\wq\) is an odd function, and thus we have \(\wp(\half\prd)
= \wp(-\half\prd) + c\) by periodicity and \(\wp(\half\prd) = \wp(-\half\prd)\)
by symmetry, and hence \(c=0\). We  have thus shown that \[\wp(z) \defn {1\over
z^2}  + \int_0^z d\zeta  \left\{\wq(\zeta) +  {2\over\zeta^3} \right\}\]  is an
elliptic function. Its  only singularities are a double pole  at the origin and
its periodic images.

If we expand \(\wp\) in a Laurent  series about the origin we obtain \[\wp(z) =
{1\over   z^2}  +   \sum_{j=1}^\infty   \sum_{m,m'\in\Z  \atop   |m|+|m'|\neq0}
{(2j+1)z^{2j}\over   (m\prd  +   m'\prd')^{2(j+1)}}  \equiv   {1\over   z^2}  +
{g_2\over20} z^2  + {g_3\over  28} z^4 +  \cdots,\] where the  coefficients are
functions only of the periods
\begin{equation}
{g_2\over60}   =  \sum_{m,m'\in\Z   \atop  |m|+|m'|\neq0}   {1\over   (m\prd  +
m'\prd')^4},  \quad\textrm{and}\quad   {g_3\over140}  =  \sum_{m,m'\in\Z  \atop
|m|+|m'|\neq0} {1\over (m\prd + m'\prd')^6}.\label{eq:A}
\end{equation}
From   this  we   find   \(\wp'(z)   =  -2   z^{-3}   +  \rational1{10}g_2z   +
\rational17g_3z^3 + \cdots\),  and therefore \(\left[\wp'(z)\right]^2 = 4z^{-6}
\left\{ 1  - \rational1{10}g_2z^4 -  \rational17g_3z^6 + \cdots  \right\}\) and
\(\left[\wp(z)  \right]^3   =  z^{-6}   \left\{  1  +   \rational3{20}g_2z^4  +
\rational3{28}g_3z^6  +  \cdots  \right\}\).  Putting these  together  we  find
\[[\wp'(z)]^2 -  4[\wp(z)]^3 + g_2\wp(z) = -g_3  + Az^2 + Bz^4  + \cdots.\] The
left-hand  side is  an elliptic  function with  periods \(\prd\)  and \(\prd'\)
whose only poles are at the origin and its periodic images, the right-hand side
has the value \(-g_3\) at the  origin, and thus by \liouville's theorem it must
be a constant. We thus have  \([\wp'(z)]^2 = 4[\wp(z)]^3 - g_2\wp(z) - g_3\) as
the differential equation satisfied by \(\wp\). Indeed, this equation allows us
to express all the derivatives of \(\wp\) in terms of \(\wp\) and \(\wp'\); for
example
\begin{equation}
  \begin{array}{rl@{\qquad}rl}
    \wp''  &=  6\wp^2  -  \half  g_2, & \wp'''  &=  12\wp\wp', \\
    \wp^{(4)} &= 6(20\wp^3 - 3g_2\wp - 2g_3), &
      \wp^{(5)} &= 18 (20\wp^2 - g_2) \wp'.
  \end{array}\label{eq:wpd2} 
\end{equation}

We  can formally  solve the  differential equation  for \(\wp\)  to  obtain the
\emph{elliptic integral} which is the  functional inverse of \(\wp\) (for fixed
periods  \(\prd\)  and   \(\prd'\)),  \[z  =  \int_0^z  d\zeta   =  -  \int_0^z
{d\zeta\,\wp'(\zeta)\over  \sqrt{4\wp(\zeta)^3   -  g_2\wp(\zeta)  -   g_3}}  =
\int_{\wp(z)}^\infty {dw\over\sqrt{4w^3 - g_2w - g_3}}.\]

It is  useful to factor the  cubic polynomial which occurs  in the differential
equation,  \(\wp'^2(z) =  4(\wp-e_1)(\wp-e_2)(\wp-e_3)\),  where the  symmetric
polynomials of the roots  satisfy \(e_1 + e_2 + e_3 =  0\), \(e_1e_2 + e_2e_3 +
e_3e_1 = -\quarter  g_2\), \(e_1e_2e_3 = \quarter g_3\), and  \(e_1^2 + e_2^2 +
e_3^2 = (e_1 + e_2 + e_3)^2 - 2(e_1e_2 + e_2e_3 + e_3e_1) = \half g_2\).

Since \(\wp'\) is an odd function we have \(-\wp'(\half\prd) = \wp'(-\half\prd)
= \wp'(\half\prd) = 0\), and likewise \(\wp'(\half\prd') = 0\) and \(\wp'\left(
\half(\prd+\prd')\right) = 0\). The values  of \(\wp\) at the half-periods must
be  distinct, for  if \(\wp(\half\prd)  = \wp(\half\prd')\)  then  the elliptic
function   \(\wp(z)  -   \wp(\half\prd')\)  would   have  a   double   zero  at
\(z=\half\prd'\)  and at \(z  = \half\prd\),  which would  violate \liouville's
theorem.  Since the  \(\wp'\) vanishes  at  the half  periods the  differential
equation  implies that  \(\wp\left(\half\prd\right)  = e_1\),  \(\wp\left(\half
\prd'\right) = e_2\), \(\wp\left(\half(\prd +  \prd') \right) = e_3\), and that
\(e_1\), \(e_2\), and \(e_3\) are all distinct.

The solution of the corresponding  differential equation with a generic quartic
polynomial, \(y'^2 = a (y-r_1)  (y-r_2) (y-r_3) (y-r_4)\) with \(r_i\neq r_j\),
is  easily  found  in terms  of  the  {\weierstrass}  function by  a  conformal
transformation. First one root is  mapped to infinity by the transformation \(y
=  r_4  +  1/x\),  giving  \(x'^2  =  -a  (x-\rho_1)  (x-\rho_2)  (x-\rho_3)  /
\rho_1\rho_2\rho_3\)   with   \(\rho_j  =   1/(r_j-r_4)\).   Then  the   linear
transformation \(x = A\xi + B\)  with \(A = -4 \rho_1\rho_2\rho_3/a\) and \(B =
(\rho_1 +  \rho_2 +  \rho_3)/3\) maps this  to \(\xi'^2 =  4(\xi-e_1) (\xi-e_2)
(\xi-e_3)\) where \(e_j  = (\rho_j - B)/A\).  The solution is thus \(y  = r_4 +
1/(A\wp  + B)\),  where \(\wp\)  has the  periods implicitly  specified  by the
roots~\(e_j\).

It is  not obvious that  there exist periods  \(\prd\) and \(\prd'\)  such that
\(g_2\) and  \(g_3\) are  given by (\ref{eq:A}),  nevertheless this is  so (see
\cite{achieser:1990} for a proof).

A simple  example of  this is given  by the \emph{{\jacobi}  elliptic function}
\(\sn  z\), which is  defined by  \(z \defn  \int_0^{\sn z}  dt\, \left[(1-t^2)
(1-k^2t^2)  \right]^{-\half}\), and hence  satisfies the  differential equation
\(({\sn}'  z)^2   =  \left(1+(\sn  z)^2\right)   \left(1-k^2(\sn  z)^2\right)\)
together with the  boundary condition \(\sn0=0\). We may move  one of the roots
to infinity by  substituting \(\sn z = 1 + 1/x(z)\)  and multiplying through by
\(x(z)^4\),  giving  \(x'^2  =  -2(1-k^2)  \left[x  +  \half\right]  \left[x  -
k/(1-k)\right] \left[x + k/(1+k)\right]\). The linear change of variable \(x(z)
=  - \left[12\xi(z) +  1 -  5k^2\right]/\left[6(1-k^2)\right]\) then  puts this
into \weierstrass's canonical form  \(\xi'^2 = 4(\xi-e_1) (\xi-e_2) (\xi-e_3) =
4\xi^3 - g_2\xi - g_3\) with the roots
\begin{equation}
e_1 = \frac{k^2+1}6, \quad e_{\rational{5\pm1}2} = -\frac{k^2 \pm 6k + 1}{12};
\label{eq:B}
\end{equation}
and  correspondingly \(g_2 =  \rational1{12}(k^4 +  14k^2 +  1)\), and  \(g_3 =
\rational1{6^3}(k^2  + 1)  (k^2  +  6k +  1)  (k^2 -  6k  +  1)\). Clearly  the
{\weierstrass}  function \(\wp(z)\)  with  periods corresponding  to the  roots
\(e_j\) is a solution to this  equation. A more general solution may be written
as \(\xi(z)  = \wp(f(z))\) for some analytic  function \(f\); for this  to be a
solution  it  must  satisfy  the  differential equation,  which  requires  that
\(f'(z)^2 = 1\), so \(\xi(z) = \wp(\pm z + \Delta)\) with \(\Delta\) a suitable
constant  chosen to  satisfy the  boundary conditions.  It turns  out  that the
boundary values  required for  \(\sn\) are satisfied  by the choice  \(\xi(z) =
\wp(z  -\K(k))\), where  \(\K(k) \defn  \int_0^1 dt\,  \left[(1-t^2) (1-k^2t^2)
\right]^{-\half}\)  is the  \emph{complete elliptic  integral}. We  shall later
derive  the expression  for \(\sn\)  in terms  of the  {\weierstrass} functions
\(\wp\) and \(\wp'\) with the same argument and periods by a simpler method.

\subsubsection{The {\weierstrass} \(\wzeta\)-Function}

It  is useful  to consider  integrals  of \(\wp\),  even though  these are  not
elliptic  functions.  If  we  define  \(\wzeta' =  -\wp\),  whose  solution  is
\[\wzeta(z)  \defn  {1\over z}  -  \int_0^z  du  \left\{\wp(u) -  {1\over  u^2}
\right\}\]  where the  path  of  integration avoids  all  the singularities  of
\(\wp\) (i.e., the periodic images of the origin) except for the origin itself.
The only singularities of \(\wzeta\) are a simple pole with unit residue at the
origin and  its periodic  images. Furthermore, \(\wzeta\)  is an  odd function.
However, \(\wzeta\)  is not periodic: \(\wzeta(z  + \prd) =  \wzeta(z) + \eta\)
where \(\eta\)  is a constant.  Setting \(z =  -\half\prd\) and using  the fact
that  \(\wzeta\) is odd  we obtain  \(\wzeta(-\half\prd) =  \wzeta(\half\prd) +
\eta  = -  \wzeta(\half\prd)\), or  \(\wzeta(\half\prd) =  \half \eta\).  If we
integrate \(\wzeta\) around a  period parallelogram \(P\) containing the origin
we find a useful identity relating \(\prd\), \(\prd'\), \(\eta\) and \(\eta'\):
\(2\pi  i   =  \oint_{\partial  P}   dz\,  \wzeta(z)  =   \int_c^{c+\prd}  dz\,
\left[\wzeta(z)   -   \wzeta(z   +   \prd')\right]  +   \int_c^{c+\prd'}   dz\,
\left[\wzeta(z  +  \prd) -  \wzeta(z)\right]  =  \int_c^{c+\prd'}  dz\, \eta  -
\int_c^{c+\prd} dz\, \eta' = \eta\prd' - \eta'\prd\).

\subsubsection{The {\weierstrass} \(\wsigma\)-Function}

That was so much fun that we  will do it again. Let \((\ln\wsigma)' = \wsigma'/
\wsigma  =  \wzeta\),  so  \[\wsigma(z) \defn  z\exp\left[\int_0^z  du  \left\{
\wzeta(u) -  {1\over u}  \right\} \right],\] where  again the  integration path
avoids all the singularities of  \(\wzeta\) except the origin. \(\wsigma\) is a
holomorphic  function having  only simple  zeros lying  at the  origin  and its
periodic images, and it is odd. To find the values of \(\wsigma\) on the period
lattice we integrate \(\wsigma'(z +  \prd)/\wsigma(z + \prd) = \wzeta(z + \prd)
= \wzeta(z) + \eta = \wsigma'(z)/\wsigma(z) + \eta\) to obtain \(\ln\wsigma(z +
\prd) = \ln\wsigma(z)  + \eta z + c\),  or \(\wsigma(z + \prd) =  c' e^{\eta z}
\wsigma(z)\).  As usual  we can  find the  constant \(c'\)  by  evaluating this
expression  at  \(z  =  -\half\prd\),  \(\wsigma(\half\prd) =  -  c'  e^{-\half
\eta\prd}   \wsigma(\half\prd)\),  giving   \(c'  =   -e^{\half\eta\prd}\)  and
\(\wsigma(z + \prd) = - e^{\eta(z + \half\prd)} \wsigma(z)\).

\subsection{Expansion of Elliptic Functions} \label{sec:expansion}

Every  rational function  \(R(z)\) can  be  expressed in  two canonical  forms,
either in a  fully factored representation which makes all  the poles and zeros
explicit,  \[R(z) =  c {(z-b_1)  (z-b_2) \cdots  (z-b_n) \over  (z-a_1) (z-a_2)
\cdots (z-a_m)},\] or  in a partial fraction expansion  which makes the leading
``divergent'' part of its Laurent  expansion about its poles manifest, \[R(z) =
E(z) +  \sum_{i,k} {A_k^{(i)}\over  (z-a_k)^i}.\] In these  expressions \(b_i\)
are  the  zeros  of \(R\),  \(a_i\)  its  poles,  \(c\) and  \(A_k^{(i)}\)  are
constants, and  \(E\) is a polynomial. It  is perhaps most natural  to think of
\(E\), the  \emph{entire part} of  \(R\), as the  leading terms of  its Laurent
expansion about infinity.

An arbitrary elliptic function \(f\) with periods \(\prd\) and \(\prd'\) may be
expanded in two analogous ways in terms of {\weierstrass} elliptic functions
with the same periods.

\subsubsection{Multiplicative Form}

To obtain the  first representation recall that \(\sum_{j=1}^n (a_j  - b_j) = 0
\pmod{\prd,\prd'}\), so we can choose a set of poles and zeros (not necessarily
in the  fundamental parallelogram)  whose sum is  zero. For instance,  we could
just  add  the appropriate  integer  multiples  of  \(\prd\) and  \(\prd'\)  to
\(a_1\). We now construct  the function \[g(z) = {\wsigma(z-b_1) \wsigma(z-b_2)
\cdots     \wsigma(z-b_n)\over     \wsigma(z-a_1)     \wsigma(z-a_2)     \cdots
\wsigma(z-a_n)},\] which has the same zeros and poles as \(f\). Furthermore, it
is  also  an  elliptic  function,   since  \(g(z  +  \prd)  =  \exp\left[  \eta
\sum_{j=1}^n  (a_j -  b_j) \right]  g(u) =  g(u)\). It  follows that  the ratio
\(f(u)/g(u)\) is  an elliptic function with no  poles, as for each  pole in the
numerator there is  a corresponding pole in the denominator,  and for each zero
in the denominator  there is a corresponding zero  in the numerator. Therefore,
by \liouville's theorem, the ratio must  be a constant \(f(u)/g(u) = C\), so we
have  \[f(z)  =  C  {\wsigma(z-b_1) \wsigma(z-b_2)  \cdots  \wsigma(z-b_n)\over
\wsigma(z-a_1) \wsigma(z-a_2) \cdots \wsigma(z-a_n)}.\]

\subsubsection{Additive Form}

For the  second ``partial  fraction'' representation let  \(a_1,\ldots,a_n\) be
the  poles of  \(f\) lying  in some  fundamental parallelogram.  In  this case,
unlike the previous  one, we ignore multiplicity and count  each pole just once
in this  list. Further, let  the leading terms  of the Laurent  expansion about
\(z=a_k\)  be \(\sum_{r=1}^{m_k}  (-1)^r (r-1)!  A_k^{(r-1)}(z-a_k)^{-r}\); the
function \(g_k(z) \equiv - \sum_{r=1}^{m_k} A_k^{(r-1)} \wzeta^{(r-1)}(z-a_k) =
- A_k^{(0)}  \wzeta(z-a_k) + \sum_{r=2}^{m_k}  A_k^{(r-1)} \wp^{(r-2)}(z-a_k)\)
then has exactly the same leading terms in its Laurent expansion.

Summing this  expression over  all the poles,  we obtain \(g(z)  = \sum_{k=1}^n
g_k(z)  = -\sum_{k=1}^n  \sum_{r=1}^{m_k}  A_k^{(r-1)} \wzeta^{(r-1)}(z-a_k)\).
The  sum  of the  terms  with  \(r>1\), being  sums  of  the elliptic  function
\(\wp(z-a_k)\) and its  derivatives, is an elliptic function.  The sum of terms
with \(r=1\),  \(\varphi(z) = -\sum_{k=1}^n  A_k^{(0)} \wzeta(z-a_k)\), behaves
under  translation by  a period  as  \(\varphi(z +  \prd) =  \varphi(z) -  \eta
\sum_{k=1}^n  A_k^{(0)} =  \varphi(z)\), where  we have  used the  corollary of
\liouville's  theorem that the  sum of  the residues  at all  the poles  of the
elliptic function \(f\) in a fundamental parallelogram is zero. It follows that
the sum of \(r=1\) terms is an elliptic function also, so the difference \(f(z)
- g(z)\)  is  an   elliptic  function  with  no  singularities,   and  thus  by
\liouville's theorem is a constant \(C\).  We have thus obtain the expansion of
an  arbitrary  elliptic  function  \(f(z)  =  C  +  g(z)  =  C  -  \sum_{k=1}^n
\sum_{r=1}^{m_k}  A_k^{(r-1)} \wzeta^{(r-1)}  (z-a_k)\),  where the  \(\wzeta\)
functions have the same periods as \(f\) does.

\subsubsection{Addition Theorems} \label{sec:addition}

Consider the  elliptic function \(f(u)  = \wp'(u)/\left(\wp(u)-\wp(v)\right)\);
according to \liouville's theorem the  denominator must have exactly two simple
zeros,  at which  \(\wp(u) =  \wp(v)\), within  any  fundamental parallelogram.
\(\wp\) is an even function, \(\wp(-v)  = \wp(v)\), so these zeros occur at \(u
= \pm  v\). At \(u=0\) the  function \(f\) has  a simple pole, and  the leading
terms of the Laurent series for \(f\) about these three poles is \((u-v)^{-1} +
(u+v)^{-1} - 2/u\). The ``partial  fraction'' expansion of \(f\) is thus \(f(u)
=  C +  \wzeta(u-v) +  \wzeta(u+v)  - 2\wzeta(u)\),  and since  both \(f\)  and
\(\wzeta\) are odd functions we observe that \(C=0\).

Adding  this  result,   \(\wp'(u)/\left(\wp(u)-\wp(v)\right)  =  \wzeta(u-v)  +
\wzeta(u+v) - 2\wzeta(u)\), to the  corresponding equation with \(u\) and \(v\)
interchanged, \(- \wp'(v)/\left(\wp(u)-\wp(v)\right) = - \wzeta(u-v) + \wzeta(u
+v)  -  2\wzeta(v)\), gives  \(\left(\wp'(u)  -  \wp'(v)\right) /  \left(\wp(u)
-\wp(v)\right)  = 2\wzeta(u+v)  - 2\wzeta(u)  - 2\wzeta(v)\).  Rearranging this
gives the \emph{addition theorem}  for zeta functions \(\wzeta(u+v) = \wzeta(u)
+  \wzeta(v) +  \half  \left(\wp'(u) -  \wp'(v)\right)  / \left(\wp(u)  -\wp(v)
\right)\).

The  corresponding   addition  theorem  for  \(\wp\)  is   easily  obtained  by
differentiating this relation \[-\wp(u+v) =  - \wp(u) + \half {\wp''(u) [\wp(u)
- \wp(v)] - \wp'(u) [\wp'(u) - \wp'(v)] \over [\wp(u) - \wp(v)]^2}\] and adding
to it the same formula with \(u\) and \(v\) interchanged to obtain \[-2\wp(u+v)
=  - \wp(u)  -  \wp(v)  + \half  {[\wp''(u)  - \wp''(v)]  [\wp(u)  - \wp(v)]  -
[\wp'(u)-\wp'(v)]^2\over [\wp(u)  - \wp(v)]^2}.\] Recalling  that a consequence
of    the    differential    equation    satisfied   by    \(\wp\)    is    the
identity~(\ref{eq:wpd2}), \(2\wp''  = 12\wp^2  - g_2\), we  have \(\wp''  (u) -
\wp''(v) =  6[\wp(u)^2 - \wp(v)^2]\), and  thus \(\wp(u+v) =  -\wp(u) -\wp(v) +
\quarter \left({\wp'(u) - \wp'(v)\over \wp(u) - \wp(v)} \right)^2\).

Differentiating this  addition theorem for  \(\wp\) gives the  addition theorem
for \(\wp'\). Since higher derivatives of  \(\wp\) can be expressed in terms of
\(\wp\) and \(\wp'\) there is no need to repeat this construction again.

\subsubsection{Representation of Elliptic Functions in terms of \(\wp\) and
\(\wp'\)} \label{sec:representation}

Consider the  ``partial fraction'' expansion of an  arbitrary elliptic function
\(f\)  in  terms  of  zeta  functions  and their  derivatives,  \(f(z)  =  C  +
\sum_{k=1}^n   \sum_{r=1}^{m_k}  A_k^{(r-1)}  \wzeta^{(r-1)}   (z-a_k)\).  This
expresses \(f(z)\) as a linear combination of zeta functions \(\wzeta(z-a_k)\),
which are not elliptic functions, and their derivatives \(\wzeta^{(r)}(z-a_k) =
-\wp^{(r-1)}(z-a_k)\) which are. We may  now use the addition theorems to write
this  in  terms  of  the   zeta  function  \(\wzeta(z)\)  and  its  derivatives
\(\wzeta^{(r)}(z) = -\wp^{(r-1)}(z)\) of the unshifted argument~\(z\).

For  the   \(r=1\)  terms   the  zeta  function   addition  theorem   gives  us
\(\sum_{k=1}^n  A_k^{(0)} \wzeta(z-a_k)  = \sum_{k=1}^n  A_k^{(0)}  \wzeta(z) +
R_1\left(\wp(z),  \wp'(z)\right)\), were  we use  the notation  \(R_i(x,y)\) to
denote a  rational function of \(x\) and  \(y\); i.e., an element  of the field
\(\C(x,y)\). The coefficients in \(R_1\) depend transcendentally on \(a_k\), of
course.   This   expression   simplifies   to  just   the   rational   function
\(R_1(\wp,\wp')\)   on   recalling  that,   as   we   have  previously   shown,
\(\sum_{k=1}^n A_k^{(0)} = 0\).

Using the addition theorems for \(\wp\)  and \(\wp'\) all the terms for \(r>1\)
may be expressed in the form \(\sum_{k=1}^n A_k^{(r-1)} \wzeta^{(r-1)}(z-a_k) =
- \sum_{k=1}^n   A_k^{(r-1)}  \wp^{(r-2)}(z-a_k)  =   R_r\left(\wp(z),  \wp'(z)
\right)\). We  have thus  shown that  \(f = R(\wp,\wp')\).  In fact,  since the
differential  equation for  \(\wp\)  expresses \(\wp'^2\)  as  a polynomial  in
\(\wp\) we  can simplify this  to the form  \(f = R_e(\wp) +  R_o(\wp)\wp'\). A
simple corollary is that if \(f\) is an even function then \(f = R_e(\wp)\) and
if it is odd then \(f = R_o(\wp) \wp'\).

A corollary  of this result is  that \emph{any two elliptic  functions with the
same periods are algebraic functions of each other}. If \(f\) and \(g\) are two
such functions then \(f = R_1(\wp)  + R_2(\wp)\wp'\), \(g = R_3(\wp) + R_4(\wp)
\wp'\), and \({\wp'}^2  = 4\wp^3 - g_2\wp -  g_3\); these equations immediately
give three polynomial  relations between the values \(f\),  \(g\), \(\wp\), and
\(\wp'\), and  we may eliminate  the last two  to obtain a  polynomial equation
\(F(f,g)  =  0\). To  be  concrete,  suppose  \(R_i(z) =  P_i(z)/Q_i(z)\)  with
\(P_i,Q_i \in \C[z]\), then we have
\begin{eqnarray*}
\bar  P_1(f,\wp,\wp')  &\equiv&  Q_1(\wp)  Q_2(\wp)  f -  P_1(\wp)  Q_2(\wp)  -
P_2(\wp) Q_1(\wp) \wp' = 0,  \\ \bar P_2(g,\wp,\wp') &\equiv& Q_3(\wp) Q_4(\wp)
g  - P_3(\wp)  Q_4(\wp) -  P_4(\wp) Q_3(\wp)  \wp' =  0, \\  \bar P_3(\wp,\wp')
&\equiv& {\wp'}^2 - 4\wp^3 + g_2\wp + g_3 = 0;
\end{eqnarray*}
we  may  then construct  the  resultants\footnote{In  practice Gr\"obner  basis
methods might be preferred.}
\begin{eqnarray*}
\bar   P_4(f,\wp)  &\equiv&   \res_{\wp'}\left(\bar   P_1(f,  \wp,\wp'),   \bar
P_3(\wp,\wp')\right)  = 0,  \\ \bar  P_5(g,\wp)  &\equiv& \res_{\wp'}\left(\bar
P_2(g,\wp,  \wp'),   \bar  P_3(\wp,\wp')  \right)  =  0,   \\  F(f,g)  &\equiv&
\res_{\wp}\left(\bar P_4(f,\wp), \bar P_5(g, \wp) \right) = 0.
\end{eqnarray*}

A corollary of this corollary is obtained by letting \(g = f'\), which tells us
that every elliptic  function satisfies a first order  differential equation of
the form \(F(f,f') = 0\) with \(F\in\C[f,f']\).

A second metacorollary is obtained  by considering \(g(u) = f(u+v)\), for which
we  deduce that there  is a  polynomial \(\C\langle  v\rangle[f(u)][f(u+v)] \ni
F=0\),   where  \(\C\langle   v\rangle\)   is  the   space  of   complex-valued
transcendental functions of  \(v\). On the other hand,  interchanging \(u\) and
\(v\)  we  observe  that  \(F\in\C\langle u\rangle  [f(v)][f(u+v)]\)  too.  The
coefficients  of  \(F\)  are   therefore  both  polynomials  in  \(f(u)\)  with
coefficients  which are functions  of~\(v\), and  polynomials in  \(f(v)\) with
coefficients  which  are functions  of~\(u\).  It  therefore  follows that  the
coefficients  must  be  polynomials  in  \(f(u)\) and  \(f(v)\)  with  constant
coefficients,  \(F\in\C[f(u),f(v),f(u+v)]\).  In  other words,  every  elliptic
equation has an algebraic addition theorem.

\subsection{{\jacobi} Elliptic Functions} \label{sec:jacobi}

We  shall  now consider  the  {\jacobi}  elliptic  function \(\sn\)  implicitly
defined     by      \(z     \defn     \int_0^{\sn      z}     dt\,\left[(1-t^2)
(1-k^2t^2)\right]^{-\half}\).  This  cannot be  anything  new  ---  it must  be
expressible  rationally in terms  of the  {\weierstrass} functions  \(\wp\) and
\(\wp'\) with the same periods.

The  integrand of  the  integral  defining \(\sn\)  has  a two-sheeted  Riemann
surface with four  branch points. The values of  \(z\) for which \(\sn(z,k)=s\)
for  any  particular  value  \(s\in\C\)  are  specified  by  the  integral;  we
immediately see that there two such values, corresponding to which sheet of the
integrand we  end up on,  plus arbitrary integer  multiples of the  two periods
\(\prd\) and \(\prd'\). These  periods correspond to the non-contractible loops
that encircle  any pair of  the branch points.  There are only  two independent
homotopically non-trivial  loops because the  contour which encircles  all four
branch points is contractible through the  point at infinity (this is a regular
point of the integrand, as may be seen by changing variable to \(1/z\)).

We may choose the first period  to correspond to a contour \(C\) which contains
the branch points at \(z=\pm1\). We find \[\prd = \oint_C \! {dt\over \sqrt{(1-
t^2)  (1-k^2t^2)}} =  \left[\int_0^1  \!\!\! -  \int_1^0  \!\!\! -  \int_0^{-1}
\!\!\!\!\!  +  \int_{-1}^0  \right]  {dt  \over  \sqrt{(1-t^2)  (1-k^2t^2)}}  =
4\K(k),\] taking into account the fact that the integrand changes sign as we go
round each branch  point onto the other sheet of the  square root. Likewise, we
may choose  the second period to  correspond to a contour  \(C'\) enclosing the
branch  points at \(z=1\)  and \(z=1/k\);  this gives  \[\prd' =  \oint_{C'} \!
{dt\over\sqrt{(1-t^2) (1-k^2t^2)}} =  \left[ \int_1^{1/k} \!\!\! - \int_{1/k}^1
\right] {dt\over \sqrt{(1-t^2)  (1-k^2t^2)}}.\] If we change variable  to \(s =
\sqrt{(1  -  k^2 t^2)/(1-k^2)}\)  we  find  that  \[\int_1^{1/k} \!\!  {dt\over
\sqrt{(1-t^2) (1-k^2 t^2)}} = i\int_0^1 \! {ds\over \sqrt{(1-s^2) (1-k'^2s^2)}}
= i\K(k'),\]  where we define  \(k' \defn\sqrt{1 -  k^2}\). We thus  have shown
that the  second period  \(\prd'=2i\K(k')\) is also  expressible as  a complete
elliptic integral.

The locations  of the  poles of  \(\sn\) are also  easily found.   Consider the
integral  \(\int_{1/k}^\infty  dt\,  [(1-t^2)  (1-k^2t^2)]^{-\half}\),  by  the
change  of variable  \(s=1/kt\) we  see  that it  is equal  to \(\int_0^1  ds\,
(ks^2)^{-1}   \left[\left(1-1/k^2s^2\right)   \left(1-1/s^2\right)   \right]^{-
\half} =  \K(k)\).  We therefore  have that \[\int_0^\infty  \!\!\!\!  {dt\over
\sqrt{(1-t^2) (1-k^2t^2)}} =  \left[\int_0^1 \!\!\! + \int_1^{1/k} \!\!\!\!\!\!
+  \int_{1/ k}^\infty \right]  {dt\over \sqrt{(1-t^2)  (1-k^2t^2)}} =  2\K(k) +
i\K(k'),\] and thus \(\sn\) has a pole at \(2\K(k) + i\K(k')\).

We  mentioned  that there  are  always  two  locations within  the  fundamental
parallelogram at which \(\sn(z,k)=s\). One  of these locations corresponds to a
contour \(C_1\)  which goes from \(t=0\)  on the principal  sheet (the positive
value of the square root in the  integrand) to \(t=s\) on the same sheet, while
the other  goes from \(t=0\)  on the principal  sheet to \(t=s\) on  the second
sheet. This latter  contour is homotopic to one which goes  from \(t=0\) on the
principal sheet to \(t=0\) on the  second sheet and then follows \(C_1\) but on
the second sheet. If  the value of the first integral is  \(z\), then the value
of the  second is \(2K(k)  - z\), thus  establishing the identity  \(\sn(z,k) =
\sn(2K(k)-z,k)\).

Since  the integrand  is  an even  function of  \(t\)  the integral  is an  odd
function  of  \(\sn\),  from  which   we  immediately  see  that  \(\sn(z,k)  =
-\sn(-z,k)\).

We summarise these  results by giving some of the values  of \(\sn\) within the
fundamental   parallelogram   defined   by  \(\prd=4K\)   and   \(\prd'=2iK'\):
\[\begin{array}{|@{\:\:}c@{\:\:}||*8{@{\:\:}c@{\:\:}|}}  \hline\phantom{\bigl|}
z & 0 & K & 2K & 3K  & iK' & K+iK' & 2K+iK' & 3K+iK' \\ \hline \phantom{\bigl|}
\sn(z,k)  &  0  &   1  &  0  &  -1  &  \infty  &  1/k   &  -\infty  &  -1/k  \\
\hline\end{array}\]  where  we have  used  the  notation  \(K\defn \K(k)\)  and
\(K'\defn\K(k')\).

\subsubsection{Representation of \(\sn\) in terms of \(\wp\) and \(\wp'\)}
\label{sec:repsn}

From this knowledge of the periods,  zeros, and poles of \(\sn\) we can express
it in  terms of  {\weierstrass} elliptic functions.  From (\ref{eq:B})  we know
that  the  periods   \(\prd=4K\),  \(\prd'=2iK'\),  and  \(\prd+\prd'=4K+2iK'\)
correspond   to   the   roots   \(e_1\),   \(e_2\),  and   \(e_3\);   that   is
\(\wp\left(\half\prd\right) = e_1\),  \(\wp\left(\half\prd'\right) = e_2\), and
\(\wp\left(\half(\prd+\prd')\right)  =  e_3\).  Since  \(\sn(z,k)\) is  an  odd
function of  \(z\) it must  be expressible as  \(R\left(\wp(z)\right) \wp'(z)\)
where  \(R\)  is  a  rational  function;  since it  has  simple  poles  in  the
fundamental parallelogram only at \(z = \half\prd' = iK'\) and \(z = \half(\prd
+   \prd')  =   2K+iK'\)  it   must   be  of   the  form   \(\sn(z,k)  =   \bar
R\left(\wp(z)\right) \wp'(z) / \left[  \left(\wp(z) - e_2\right) \left(\wp(z) -
e_3\right)  \right]\) with  \(\bar R\)  a  rational function.\footnote{Remember
that the  logarithmic derivative  of a function  \(d\ln f(z)/dz  = f'(z)/f(z)\)
always has  a simple pole at each  pole and zero of  \(f\).} The {\weierstrass}
function has a double pole at the origin, its derivative has a triple pole, and
the {\jacobi}  elliptic function \(\sn\)  has a simple  zero, so we  can deduce
that \(\bar R\left(\wp(z)\right)\) must be  regular and non-zero at the origin,
and hence \(\bar R\) is just a  constant. As \(\sn(z,k) = z + O(z^3)\) near the
origin this  constant is easily determined  by considering the  residues of the
poles in the  {\weierstrass} functions, and we obtain  the interesting identity
\(\sn(z,k) =  -\half \wp'(z) /  \left[\left(\wp(z) - e_2\right)  \left(\wp(z) -
e_3\right) \right]\).

\subsubsection{Representation of \(\sn^2\) in terms of \(\wp\)}

We  can  use  the  same   technique  to  express  \(\sn(z,k)^2\)  in  terms  of
{\weierstrass}  elliptic  functions.  The  differential equation  satisfied  by
\(s(z) \defn  \sn(z,k)^2\) is \(s'^2  = 4s(1-s)(1-k^2s)\), which is  reduced to
{\weierstrass} canonical form  \(\wzeta'^2 = 4\wzeta^3 - \bar  g_2\wzeta - \bar
g_3\)   with  \(\bar   g_2  =   \rational43(k^4+k^2+1)\)  and   \(\bar   g_3  =
\rational4{27}  (2k^2-1)(k^2-2)(k^2+1)\)  by  the  linear  substitution  \(s  =
\left(\wzeta + \third(1 + k^2)\right) / k^2\). The roots of this cubic form are
\(\bar  e_1   =  -\rational13(1   +  k^2)\),  \(\bar   e_{\rational{5\pm1}2}  =
\rational13(2 \pm k^2)\). The general  solution of this equation is \(\wzeta(z)
= \wp(\pm z  + c)\) where \(c\)  is a constant, and since  \(\sn(z,k)^2\) has a
double  pole at  \(z  = i\K(k')\)  we  have \(\sn(z,k)^2  = \left[\wp\left(z  -
i\K(k')\right) + \third(1+k^2)\right] / k^2\). This can be simplified using the
addition  formula for  {\weierstrass} elliptic  functions  to give\footnote{The
{\weierstrass} functions in this  expression implicitly correspond to the roots
\(\bar  e_i\), whereas  as those  in the  previous expression  for \(\sn(z,k)\)
corresponded   to  the   \(e_i\).}   \(\sn(z,k)^2  =   1/\left[\wp(z)  -   \bar
e_1\right]\). Of  course, we  could have seen  this immediately by  noting that
\(\sn(z,k)^2\)  is  an  even  elliptic  function with  periods  \(2\K(k)\)  and
\(2i\K(k')\) corresponding to  the roots \(\bar e_i\), and  therefore must be a
rational function of \(\wp(z)\). Since it  has a double pole at \(z = i\K(k')\)
and  a double  zero at  \(z=0\)  in whose  neighbourhood \(\sn(z,k)^2  = z^2  +
O(z^4)\) the preceding expression is uniquely determined.

A useful  corollary of this  result is that  we can express  the {\weierstrass}
function  \(\wp(z)\) with  periods  \(2\K(k)\) and  \(2i\K'(k)\) rationally  in
terms of \(\sn(z,k)^2\), namely \(\wp(z)  = \sn(z,k)^{-2} +\bar e_1\), and thus
any even  elliptic function  with these  periods may be  written as  a rational
function of \(\sn(z,k)^2\).

\subsubsection{Addition Formula for {\jacobi} Elliptic Functions}

We may  derive the explicit  addition formula for {\jacobi}  elliptic functions
using    a   method    introduced    by   Euler.    Consider   the    functions
\(s_1\defn\sn(u,k)\),  \(s_2\defn\sn(v,k)\)  where   we  shall  hold  \(u+v=c\)
constant. The  differential equations  for \(s_1\) and  \(s_2\) are  \(s_1'^2 =
(1-s_1^2)  (1-k^2s_1^2)\), \(s_2'^2  = (1-s_2^2)(1-k^2s_2^2)\),  where  we have
used a prime  to indicate differentiation with respect to  \(u\) and noted that
\(v'=-1\). Multiplying  the equations  by \(s_2^2\) and  \(s_1^2\) respectively
and   subtracting   them   gives   \(W(s_1,s_2)\cdot(s_1s_2)'  =   (s_1s_2'   -
s_2s_1')(s_1s_2' +  s_2s_1') = (s_1^2  - s_2^2)(1 - k^2s_1^2s_2^2)\),  where we
have introduced the Wronskian  \(W(s_1,s_2) \defn \det \left( \begin{array}{cc}
s_1&s_2\\ s_1'&s_2'\end{array} \right)\).  If we differentiate the differential
equations  for  \(s_1\)   and  \(s_2\)  we  obtain  \(s_1''   =  -(1+k^2)s_1  +
2k^2s_1^3\), \(s_2''  = -(1+k^2)s_2 + 2k^3s_2^3\);  subtracting these equations
gives  \(W'  =  (s_1s_2'  -  s_2s_1')'  = s_1s_2''  -  s_2s_1''  =  -2k^2s_1s_2
(s_1^2-s_2^2)  = (s_1^2-s_2^2) (1  - k^2  s_1^2 s_2^2)'  / (s_1s_2)'\).  We may
combine the expressions  we have derived for \(W\) and  \(W'\) to obtain \((\ln
W)'     =     W'/W    =     (1-k^2s_1^2s_2^2)'     /    (1-k^2s_1^2s_2^2)     =
\left(\ln(1-k^2s_1^2s_2^2)\right)'\). Upon integration  this yields an explicit
expression  for the Wronskian,  \(W =  C (1-k^2s_1^2s_2^2)\)  where \(C\)  is a
constant, by  which we mean  that it does  not depend upon \(u\).  The constant
does  depend on  the value  of \(c=u+v\),  and it  may be  found  by evaluating
formula at \(v =0\).

To do so  it is convenient to introduce two  other {\jacobi} elliptic functions
\(\cn(u,k) \defn \sqrt{1 - \sn(u,k)^2}\) where \(\cn(0,k) = 1\); and \(\dn(u,k)
\defn \sqrt{1  - k^2\sn(u,k)^2}\),  where \(\dn(0,k) =  1\). In terms  of these
functions we  may write \(\sn'u  = \cn u\dn  u\), furthermore they  satisfy the
identities \((\sn u)^2 + (\cn u)^2 =  1\) and \((k^2\sn u)^2 + (\dn u)^2 = 1\),
and differentiating these identities yields  \(\cn'u = -\sn u\dn u\), \(\dn'u =
-k^2\sn u\cn u\).

We may now write  \(C = W(\sn u,\sn v) / \left[1-(k\sn  u\sn v)^2\right] = [\sn
u\cn v\dn v + \sn v\cn  u\dn u] / \left[1-(k\sn u\sn v)^2\right]\), remembering
that \(v'=-1\). Setting \(v=0\)  gives \(C = \sn u = \sn  c\), and thus we have
the desired addition formula \(\sn(u+v) = \left(\sn u\cn v\dn v + \sn v\cn u\dn
u\right) / \left(1-(k\sn u\sn v)^2\right)\).

\subsection{Transformations of Elliptic Functions} \label{sec:modular}

So far we  have studied the dependence of elliptic  functions on their argument
for fixed values of the periods. Although the {\weierstrass} function appear to
depend on two arbitrary complex periods \(\prd\) and \(\prd'\) they really only
depend on the ratio \(\tau=\prd'/\prd\).  If we rewrite the identity \(\wp(z) =
\wp(z+\prd)  = \wp(z+\prd')\)  in  terms  of the  new  variable \(\wzeta  \defn
z/\prd\)   we  have   \(\wp(\prd\wzeta)   =  \wp\left(\prd(\wzeta+1)\right)   =
\wp\left(\prd(\wzeta +  \tau)\right)\). Viewed as  a function of  \(\wzeta\) we
have an elliptic function with periods \(1\) and \(\tau\), and as we have shown
this is expressible rationally in  terms of the {\weierstrass} function and its
derivative with these periods.

Another  observation is that  there are  many choices  of periods  \(\prd\) and
\(\prd'\) which generate the same  period lattice. Indeed, if we choose periods
\(\tilde\prd  =  \alpha\prd  +   \beta\prd'\),  \(\tilde\prd'  =  \gamma\prd  +
\delta\prd'\) with \(\det  \left( \begin{array}{cc} \alpha & \beta  \\ \gamma &
\delta\end{array}\right)  = 1\)  then this  will be  the case.  This  induces a
relation  between elliptic functions  with these  periods called  a \emph{first
degree transformation}.

\subsubsection{{\jacobi} Imaginary Transformation}

{\jacobi}'s  imaginary  transformation,  or the  second  principal\footnote{The
first  principal first  degree  transformation may  be  derived similarly.  See
\cite{achieser:1990} for details.}  first degree transformation, corresponds to
the interchange of periods \(\prd'  = -\tilde\prd\) and \(\prd = \tilde\prd'\).
We start with the function \(\sn(z,k)^2\) which has periods \(\tilde\prd = 2K\)
and \(\tilde\prd'  = 2iK'\),  and consider the  function \(\sn(z/M,\lambda)^2\)
with periods \(\prd = 2ML\) and \(\prd' = 2iML'\) (with \(L = \K(\lambda)\) and
\(L' =  \K(\lambda')\) as  usual). For suitable  \(M\) and \(\lambda\)  we have
\(ML=iK'\)  and  \(iML'=-K\),  corresponding  to  the  desired  interchange  of
periods.

Since \(\sn(z/M,\lambda)^2\)  is an even  function whose period lattice  is the
same as that of \(\sn(z,k)^2\) it must be expressible as a rational function of
\(\sn(z,k)^2\),  and  this rational  function  may  be  found by  matching  the
location  of poles  and  zeros.  \(\sn(z/M,\lambda)^2\) has  a  double zero  at
\(z/M=0\) and  a double  pole at \(z/M  = iL'\).  This latter condition  may be
written as \(z=iML'= -K\), or  \(z=K\) upon using the periodicity conditions to
map  the  pole  into  the fundamental  parallelogram.  Thus  \[\sn\left({z\over
M},\lambda\right)^2  = {A\sn(z,k)^2\over\sn(z,k)^2- \sn(K,k)^2}  = {A\sn(z,k)^2
\over \sn(z,k)^2-1}.\]

The constant  \(A\) may be found by  evaluating both sides of  this equation at
\(z  = iK'\):  on the  left  \(\sn(iK'/M,\lambda)^2 =  \sn(L,\lambda)^2 =  1\),
whereas  on the  right we  have \(A\)  because \(\sn(z,k)\to\infty\)  as \(z\to
iK'\). We thus have \(A=1\).

The value of \(\lambda\) is found  by evaluating both sides at \(z=-K+iK'\): on
the   left  \(\sn\left((-K+iK')/M,\lambda\right)^2  =   \sn(iL'+L,\lambda)^2  =
1/\lambda^2\), and on the right we have \(A/(1-k^2)\) since \(\sn(-K+iK',k)^2 =
1/k^2\). We thus have \(\lambda = \sqrt{1-k^2} = k'\).

From these values for \(A\) and  \(\lambda\) we may easily find \(M\), as \(iK'
= ML = M\K(\lambda) = M\K(k') = MK'\) gives \(M=i\). We may therefore write the
{\jacobi}  imaginary transformation  as \(\sn(-iz,k')^2  = \sn(z,k)^2  / \left(
\sn(z,k)^2 -1 \right)\), or equivalently \(\sn(iz,k') = i\sn(z,k) / \cn(z,k)\),
where we  have made use  of the  fact that \(\sn^2\)  is an even  function, and
chosen the sign  of the square root according to the  definition of \(\cn\) and
the fact that \(\sn(z,k) = z + O(z^3)\).

\subsubsection{Principal Transformation of Degree \(n\)} \label{sec:principal}

We  can also  choose periods  \(\tilde\prd\) and  \(\tilde\prd'\)  whose period
lattice  has the  original one  as  a sublattice,  for instance  we may  choose
\(\tilde\prd = \prd\) and  \(\tilde\prd' = \prd'/n\) where \(n\in\N\). Elliptic
functions with  these periods  must be rationally  expressible in terms  of the
{\weierstrass}  elliptic  functions with  the  original  periods, although  the
inverse may not be true.  This relationship is called a \emph{transformation of
degree \(n\)}.

Let us construct such an  elliptic function with periods \(4K\) and \(2iK'/n\),
where  \(K \defn\K(k)\) and  \(K'\defn\K(k')\) with  \(k^2+k'^2=1\). We  may do
this  by taking  \(\sn(z,k)\)  and scaling  \(z\)  by some  factor \(1/M\)  and
choosing a new parameter \(\lambda\). We  are thus led to consider the function
\(\sn(z/M,\lambda)\),  whose periods  with respect  to \(z/M\)  are  \(4L \defn
4\K(\lambda)\) and \(2iL' \defn 2i\K(\lambda')\), with \(\lambda^2 + \lambda'^2
= 1\). Viewed as a function of  \(z\) it has periods \(4LM\) and \(2iL'M\), and
\(M\) and  \(\lambda\) are  thus fixed by  the conditions  that \(LM =  K\) and
\(L'M  =  K'/n\). The  ratio  \(f(z)  \defn  \sn(z/M, \lambda)/\sn(z,k)\)  must
therefore be an even function of \(z\) with periods \(2K\) and \(2K'\);
\begin{eqnarray*}
\lefteqn{f(\pm z  + 2mK +  2im'K') = {\sn\left({\pm  z + 2mK +  2im'K'\over M},
\lambda\right)\over\sn(\pm  z   +  2mK  +   2im'K',k)}}  &&  \\  &&   \qquad  =
{\sn\left(\pm {z\over M} + 2mL  + 2im'nL', \lambda\right)\over\sn(\pm z + 2mK +
2im'K',k)}  =   {\pm  (-)^m  \sn\left({z\over  M},   \lambda  \right)\over  \pm
(-)^m\sn(z,k)} = f(z)
\end{eqnarray*}
for  \(m,m'\in\Z\);  and  hence  \(f(z)\)   must  be  a  rational  function  of
\(\sn(z,k)^2\).

Within its fundamental parallelogram the numerator of \(f(z)\) has simple zeros
at  \(z =  m2iL'M =  m2iK'/n\) for  \(m=0,1,\ldots,n-1\) and  simple  poles for
\(m=\half, \rational32, \ldots,n-\half\); whereas  its denominator has a simple
zero  at \(z=0\)  and  a simple  pole at  \(z=iK'\).  Hence, if  \(n\) is  even
\(f(z)\) has simple  zeros for \(m=1,2,\ldots,\half n-1,\half n+1,\ldots,n-1\),
a    double    zero    for    \(m=\half    n\),   and    simple    poles    for
\(m=\half,\rational32,\ldots,n-\half\);   whereas   if   \(n\)  is   odd   then
\(m=1,2,\ldots,n-1\) give simple  zeros and \(m=\half, \rational32,\ldots,\half
n-1,\half  n+1,\ldots,n-\half\) simple  poles. Therefore  there  are \(2\lfloor
\half n\rfloor\) zeros and  poles, and it is easy to see  that they always come
in pairs  such that the zeros  occur for \(\sn(z,k) =  \pm \sn(2iK'm/n,k)\) and
the  poles  for  \(\sn(z,k)  = \pm  \sn\left(2iK'(m-\half)/n,k  \right)\)  with
\(m=1,\ldots,\lfloor\half   n\rfloor\).   We  thus   see   that  the   rational
representation is
\begin{equation}
f(z)  \defn  {\sn\left({z\over M},\lambda\right)\over  \sn(z,k)}  = {1\over  M}
\prod_{m=1}^{\lfloor\half   n\rfloor}   {1   -  {\sn(z,k)^2   \over\sn(2iK'm/n,
k)^2}\over 1 - {\sn(z,k)^2\over\sn\left(2iK'(m-\half)/n,k \right)^2}},
\label{eq:C}
\end{equation}
where  the overall  factor  is  determined by  considering  the behaviour  near
\(z=0\).

The value of the quantity \(M\) may be determined by evaluating this expression
at the half period \(K\) where \(\sn(K,k) = 1\) and \(\sn(K/M,\lambda) = \sn(L,
\lambda)  = 1\),  so \[M  =  \prod_{m=1}^{\lfloor\half n\rfloor}  {1 -  {1\over
\sn(2iK'm/n, k)^2}\over 1 - {1\over\sn\left(2iK'(m-\half)/n,k\right)^2}}.\] The
value of the parameter \(\lambda\) is found by evaluating the identity at \(z =
K   +   iK'/n\),    where   \(\sn\left({K+iK'/n\over   M},   \lambda\right)   =
\sn\left({K\over  M} + i{K'\over  nM},\lambda\right) =  \sn(L +  iL',\lambda) =
{1\over\lambda}\).

It will prove useful to write the identity in parametric form
\begin{equation}
\sn\left({z\over  M},\lambda\right)  =  {\xi\over M}  \prod_{m=1}^{\lfloor\half
n\rfloor} {1 - c_m\xi^2\over1 - c'_m\xi^2}, \label{eq:D}
\end{equation}
with  \(\xi \defn  \sn(z,k)\), \(c_m  \defn \sn(2iK'm/n,  k)^{-2}\)  and \(c'_m
\defn \sn\left(2iK'(m- \half)/n,k \right)^{-2}\). This emphasises the fact that
\(\sn(z/M,\lambda)\)  is   a  rational  function  of   \(\sn(z,k)\)  of  degree
\(\left(2\lfloor\half n\rfloor + 1, 2\lfloor\half n\rfloor\right)\).

\section{\zolotarev's Problem} \label{sec:zolotarev}

{\zolotarev}'s   fourth  problem  is   to  find   the  best   uniform  rational
approximation  to  \(\sgn  x  \defn  \vartheta(x) -  \vartheta(-x)\)  over  the
interval  \([-1,-\varepsilon]  \union [\varepsilon,1]\).  This  is easily  done
using the identity (\ref{eq:D}) derived in the preceding section.

We  note that  the function  \(\xi =  \sn(z,k)\)  with \(k  < 1\)  is real  and
increases  monotonically in  \([0,1]\) for  \(z\in[0,K]\), where  as  before we
define \(K\defn\K(k)\) to be a complete elliptic integral. Similarly we observe
that  \(\sn(z,k)\)  is real  and  increases  monotonically  in \([1,1/k]\)  for
\(z=K+iy\)  with \(y\in[0,K']\) and  \(K'\defn\K(k')\), \(k^2+k'^2=1\).  On the
other  hand,   \(\sn(z/M,\lambda)\)  has  the   same  real  period   \(2K\)  as
\(\sn(z,k)\)  and has  an imaginary  period  \(2iK'/n\) which  divides that  of
\(\sn(z,k)\)  exactly \(n\)  times. This  means that  \(\sn(z/M,\lambda)\) also
increases monotonically in \([0,1]\)  for \(z\in[0,K]\), and then oscillates in
\([1,1/\lambda]\) for \(z=K+iy\) with \(y\in[0,K']\).

In  order to produce  an approximation  of the  required type  we just  need to
rescale both the argument \(\xi\) so  it ranges between \(-1\) and \(1\) rather
than \(-1/k\) and \(1/k\), and the function so that it oscillates symmetrically
about \(1\) for \(\xi\in[1,1/k]\)  rather than between \(1\) and \(1/\lambda\).
We thus obtain
\begin{equation}
R(x)   =   {2\over1+{1\over\lambda}}   {x\over  kM}   \prod_{m=1}^{\lfloor\half
n\rfloor} {k^2 - c_mx^2\over k^2 - c'_mx^2} \label{eq:E}
\end{equation}
with   \(k  =   \varepsilon\).   On   the  domain   \([-1,-\varepsilon]  \union
[\varepsilon,1]\) the  error \(e(x) \defn  R(x) - \sgn(x)\)  satisfies \(|e(x)|
\leq \Delta  \defn {1 - \lambda\over1 +  \lambda}\), or in other  words \(\|R -
\sgn\|_\infty =  \Delta\). Furthermore,  the error alternates  \(4\lfloor \half
n\rfloor  +  2\) times  between  the extreme  values  of  \(\pm\Delta\), so  by
\chebyshev's  theorem  on optimal  rational  approximation  \(R\)  is the  best
rational  approximation of  degree  \((2\lfloor\half n\rfloor+1,  2\lfloor\half
n\rfloor)\).   In  fact we  observe  that  \(R\)  is \emph{deficient},  as  its
denominator  is of  degree one  lower than  this;  this must  be so  as we  are
approximating  an odd  function. Indeed,  we may  note that  \(R'(x)  \defn (1-
\Delta^2) / R(x)\) is also an optimal rational approximation.

\section{Numerical Evaluation of Elliptic Functions} \label{sec:numeval}

We wish  to consider \gauss' arithmetico-geometric  mean as it  provides a good
means of evaluating {\jacobi} elliptic functions numerically.

\subsection{{\gauss} Transformation}

{\gauss} considered  the transformation  that divides the  second period  of an
elliptic function by  two, \(\prd'_1 = \prd_1\) and  \(\prd'_2 = \half\prd_2\).
This is a special case of the principal transformation of \(n\)th degree on the
second   period   considered    before   (\ref{eq:C})   with   \(n=2\),   hence
\[\sn\left({z\over   M},   \lambda\right)   =   {\sn(z,k)\over   M}   \left[{1-
{\sn(z,k)^2\over \sn(iK',k)^2}  \over1 - {\sn(z,k)^2\over  \sn(\half iK',k)^2}}
\right],\] with  the parameter  \(\lambda\) corresponding to  periods \(L=K/M\)
and  \(L'=K'/2M\).  Using  {\jacobi}'s  imaginary  transformation  (the  second
principal   first   degree   transformation,   with   \(\prd'_1=-\prd_2\)   and
\(\prd'_2=\prd_1\)),
\begin{equation}
\sn(iz,k) = i{\sn(z,k')\over\cn(z,k')}, \label{eq:F}
\end{equation}
we  get \[\sn\left({z\over M},\lambda\right)  = {\sn(z,k)  \over M}  \left[{1 +
{\sn(z,k)^2\cn(K',k')^2  \over \sn(K',k')^2}  \over 1  +  {\sn(z,k)^2 \cn(\half
K',k')^2  \over  \sn(\half  K',  k')^2}}\right].\] Since  \(\sn(K',k')  =  1\),
\(\cn(K',k')  = 0\),\footnote{Let  \(x  \equiv \sn(\half  K,k)\),  then by  the
addition  formula  for  {\jacobi}  elliptic  functions  \(\sn(K,k)  =  1  =  2x
\sqrt{1-x^2}   \sqrt{1-k^2x^2}   /   (1-k^2x^4)\).   Hence   \((1-k^2x^4)^2   =
4x^2(1-x^2)(1-k^2x^2)\), so \(k^4x^8  - 4k^2x^6 + 2(2+k^2)x^4 - 4x^2  + 1 = 0\)
or,  with \(z  \equiv  1/x^2 -  1\),  \([z^2 -  (1-k^2)]^2 =  0\).  Thus \(z  =
\pm\sqrt{1-k^2} =  \pm k'\),  or \(x =  1/\sqrt{1\pm k'}\). Since  \(0<x<1\) we
must   choose  the   positive   sign,  so   \(\sn\left(\half  K(k),k\right)   =
1/\sqrt{1+k'}\).} \(\sn(\half K',k') = 1/\sqrt{1+k}\), and \(\cn(\half K',k') =
\sqrt{k/(1+k)}\), we  obtain \(\sn(z/M, \lambda)  = \sn(z,k) \left[1  + k\sn(z,
k)^2 \right]^{-1}/M\).

To determine \(M\) we set \(z=K\): \(1 = \sn(K/M, \lambda) = \sn(K,k) \left[1 +
k\sn(K,k)^2\right]^{-1}\!\!\!/ M\)  \(= [1+k]^{-1}/M\), hence  \(M = 1/(1+k)\).
To  determine \(\lambda\)  we  set  \(u =  K+iK'/2\):  \[\sn\left({K\over M}  +
{iK'\over2M}, \lambda\right) = {\sn(K+\half iK',k)\over M[1 + k\sn(K+\half iK',
k)^2]}.\]  Now,  from  the  addition  formula\footnote{We  shall  suppress  the
parameter  \(k\) when it  is the  same for  all the  functions occurring  in an
expression.} \(\sn(u+v) = (\sn u\cn v\dn v + \sn v\cn u\dn u) / [1 - (k\sn u\sn
v)^2]\),  we  deduce   that  \[\sn  \left(K  +  {iK'\over2}\right)   =  {\sn  K
\cn{iK'\over2}  \dn{iK'\over2} +  \sn{iK'\over2} \cn  K\dn  K\over1 -\left(k\sn
K\sn{iK'\over2}\right)^2}  =  {\cn{iK'\over2}  \dn{iK'\over2}\over1  -  \left(k
\sn{iK' \over2}\right)^2}.\] Furthermore \(\sn(\half iK',k) = i\sn(\half K',k')
/ \cn(\half K',k')  = i / \sqrt k\), and correspondingly  \(\cn(\half iK', k) =
\sqrt{(1+k)/k}\),  and \(\dn(  \half iK',k)  = \sqrt{1+k}\),  giving  \(\sn(K +
\half iK',k)  = 1/\sqrt  k\). We thus  find \(1/\lambda =  \sn(L+iL',\lambda) =
1/2M\sqrt k\)  or \(\lambda = 2M\sqrt k  = 2\sqrt k /  (1+k)\). Combining these
results we obtain an explicit expression for \gauss' transformation \[\sn\left(
(1+k)z,{2\sqrt k\over1+k}\right) = {(1+k)\sn(z,k)\over1+k \sn(z,k)^2}.\]

\subsection{Arithmetico-Geometric Mean} \label{sec:agmean}

Let  \(a_n,b_n\in\R\)  with  \(a_n>b_n>0\),  and define  their  arithmetic  and
geometric  means  to  be  \(a_{n+1}  \defn  \half(a_n+b_n)\),  \(b_{n+1}  \defn
\sqrt{a_n b_n}\).  Since these are means we  easily see that \(a_n  > a_{n+1} >
b_n\)  and \(a_n  >  b_{n+1} >  b_n\);  furthermore \(a_{n+1}^2  - b_{n+1}^2  =
\quarter(a_n^2 + 2a_nb_n + b_n^2) -  a_nb_n = \quarter(a_n^2 - 2a_nb_n + b_n^2)
= \quarter(a_n-b_n)^2  > 0\), so  \(a_n > a_{n+1}  > b_{n+1} > b_n\).  Thus the
sequence  converges  to  the  \emph{arithmetico-geometric  mean}  \(a_\infty  =
b_\infty\).

If we choose \(a_n\) and \(b_n\) such that \(k = (a_n-b_n) / (a_n+b_n)\), e.g.,
\(a_n=1+k\) and \(b_n = 1-k\), then
\begin{eqnarray*}
1+k  &=& {(a_n+b_n)+(a_n-b_n)\over  a_n+b_n}  = {a_n\over  a_{n+1}},\\ k^2  &=&
\left({a_n-b_n\over      a_n+b_n}      \right)^2      =     {(a_n+b_n)^2      -
4a_nb_n\over(a_n+b_n)^2} =  1 - {b_{n+1}^2\over  a_{n+1}^2},\\ {4k\over(1+k)^2}
&=&  {(1+k)^2  -  (1-k)^2\over(1+k)^2}  =  1  -  \left({1-k\over1+k}\right)^2\\
&&\qquad= 1 - \left({(a_n+b_n)-(a_n-b_n)\over (a_n+b_n)+(a_n-b_n)}\right)^2 = 1
- {b_n^2\over a_n^2}.
\end{eqnarray*}
If  we   define  \(s_n  \defn\sn\left((1+k)z,{2\sqrt   k\over1+k}\right)\)  and
\(s_{n+1} \defn  \sn(z,k)\) then \gauss'  transformation tells us that  \[s_n =
{(1+k)s_{n+1}\over1+   ks_{n+1}^2}    =   {a_ns_{n+1}\over   a_{n+1}\left[1   +
\left({a_n-b_n\over       a_n+b_n}       \right)      s_{n+1}^2\right]}       =
{2a_ns_{n+1}\over(a_n+b_n)  + (a_n-b_n)s_{n+1}^2}.\]  On the  other hand  \[z =
\int\limits_0^{s_{n+1}}   {dt\over\sqrt{(1-t^2)   (1-k^2t^2)}}  =   {1\over1+k}
\int\limits_0^{s_n}    {dt\over\sqrt{(1-t^2)    \left[1-   {4k\over    (1+k)^2}
t^2\right]}},\]  and   these  two   integrals  may  be   rewritten  as   \[z  =
\int\limits_0^{s_{n+1}} {dt\over\sqrt{(1-t^2)\left[1 - \left(1-{b_{n+1}^2 \over
a_{n+1}^2}\right)t^2\right]}}   =    {a_{n+1}\over   a_n}   \int\limits_0^{s_n}
{dt\over\sqrt{(1-t^2)\left[1-\left(1 - {b_n^2\over a_n^2}\right)t^2\right]}}.\]
Therefore  the quantity  \[{z\over a_{n+1}}  =  \int_0^{s_{n+1}} \!\!\!\!\!\!\!
{dt\over\sqrt{(1  -t^2)  [a_{n+1}^2 (1-t^2)  +  b_{n+1}^2t^2]}} =  \int_0^{s_n}
{dt\over\sqrt{(1-t^2)  [a_n^2  (1-t^2) +  b_nt^2]}}\]  is  invariant under  the
transformation  \((a_n, b_n,  s_n) \mapsto  (a_{n+1}, b_{n+1},  s_{n+1})\), and
thus     \[{z\over      a_{n+1}}     =     \int_0^{s_\infty}     {dt\over\sqrt{
(1-t^2)\left[a_\infty^2(1-t^2)  + b_\infty^2t^2  \right]}} =  {1\over a_\infty}
\int_0^{s_\infty} {dt\over\sqrt{1-t^2}} = {\sin^{-1} s_\infty\over a_\infty}.\]
This implies  that \(s_\infty =  \sin\left( {a_\infty z\over  a_{n+1}}\right) =
\sin(a_\infty  z)\) with  our previous  choice of  \(a_n=1+k,  b_n=1-k \implies
a_{n+1}=1, b_{n+1}=\sqrt{1-k^2}\).  We may  thus compute \(s_{n+1} = \sn(z,k) =
f(z,1,\sqrt{1-k^2})\) for \(0<k<1\) where
\[f(z,a,b)  \defn \left\{  \begin{array}{rl}  \sin(az) &  \mbox{if \(a=b\)}  \\
{2a\xi\over (a+b) + (a-b)\xi^2}  & \mbox{with \(\xi \defn f\left(z,{a+b\over2},
\sqrt{ab} \right)\)  if \(a\neq b\).} \end{array} \right.\]  Furthermore, if we
take  \(z=K(k)\) then \(s_{n+1}=1\)  and \(s_n  = 2a_n  / [(a_n+b_n)+(a_n-b_n)]
=1\); thus  \(s_\infty = \sin(a_\infty K) =  1\), so \(a_\infty K  = \pi/2\) or
\(K(k) = \pi/2 a_\infty\).

\subsection{Computer Implementation} \label{sec:implement}

An  implementation  of  this   method  is  shown  in  Figures~\ref{fig:sncndnK}
and~\ref{fig:arithgeom}.

The  function  \texttt{arithgeom}  recursively  evaluates  the  function  \(f\)
defined above. One  subtlety is the stopping criterion, which  has to be chosen
carefully to guarantee that the recursion will terminate (which does not happen
if the simpler criterion \texttt{b==a}  is used instead) and which ensures that
the  solution is  as accurate  as  possible whatever  floating point  precision
\texttt{FLOAT}  is  specified.   Another  subtlety  is how  the  value  of  the
arithmetico-geometric mean  \texttt{*agm} is returned from  the innermost level
of the recursion. Ideally, we would like this value to be bound to an automatic
variable in  the calling  procedure \texttt{sncndnK} rather  than passed  as an
argument, thus avoiding copying its address for every level of recursion (as is
done  in here)  or copying  its value  for every  level if  it  were explicitly
returned as  a value.  Unfortunately  this is impossible, since  the \texttt{C}
programming language does not allow us to have nested procedures. The reason we
have written it in  the present form is so that the  code is thread-safe: if we
made  \texttt{agm} a  static global  variable then  two  threads simultaneously
invoking \texttt{sncndnK}  might interfere with each other's  value. The virtue
of  this approach  is  only slightly  tarnished  by the  fact  that the  global
variable \texttt{pb} used in the  convergence test is likewise not thread-safe.
The envelope  routine \texttt{sncndnK} is  almost trivial, except that  care is
needed to get the sign of \(\cn(z,k)\) correct.

\begin{figure}
\begin{verbatim}
#include <math.h>
#define ONE ((FLOAT) 1)
#define TWO ((FLOAT) 2)
#define HALF (ONE/TWO)

static void sncndnK(FLOAT z, FLOAT k, FLOAT* sn, FLOAT* cn,
                    FLOAT* dn, FLOAT* K) {
  FLOAT agm;
  int sgn;
  *sn = arithgeom(z, ONE, sqrt(ONE - k*k), &agm);
  *K = M_PI / (TWO * agm);
  sgn = ((int) (fabs(z) / *K)) % 4; /* sgn = 0, 1, 2, 3 */
  sgn ^= sgn >> 1;    /* (sgn & 1) = 0, 1, 1, 0 */
  sgn = 1 - ((sgn & 1) << 1);	/* sgn = 1, -1, -1, 1 */
  *cn = ((FLOAT) sgn) * sqrt(ONE - *sn * *sn);
  *dn = sqrt(ONE - k*k* *sn * *sn);
}
\end{verbatim}
\caption{The  procedure \texttt{sncndnK}  computes  \(\sn(z,k)\), \(\cn(z,k)\),
\(\dn(z,k)\),  and \(\K(k)\).  It  is  essentially a  wrapper  for the  routine
\texttt{arithgeom}   shown   in   Figure~\ref{fig:arithgeom}.   The   sign   of
\(\cn(z,k)\)  is defined to  be \(-1\)  if \(\K(k)  < z  < 3\K(k)\)  and \(+1\)
otherwise, and this sign is computed by some quite unnecessarily obfuscated bit
manipulations.}
\label{fig:sncndnK}
\end{figure}

\begin{figure}
\begin{verbatim}
static FLOAT arithgeom(FLOAT z, FLOAT a, FLOAT b, FLOAT* agm) {
  static FLOAT pb = -ONE;
  FLOAT xi;

  if (b <= pb) { pb = -ONE; *agm = a; return sin(z * a); }
  pb = b;
  xi = arithgeom(z, HALF*(a+b), sqrt(a*b), agm);
  return 2*a*xi / ((a+b) + (a-b)*xi*xi);
}
\end{verbatim}
\caption{Recursive implementation of  \gauss' arithmetico-geometric mean, which
is the  kernel of the method  used to compute the  {\jacobi} elliptic functions
with parameter \(k\) where \(0 < k  < 1\). The function returns a value related
to  \(\sn(z,k')\),   and  also   sets  the  value   of  \texttt{*agm}   to  the
arithmetico-geometric mean.  This value is simply related  to complete elliptic
function  \(\K(k')\)  and  also  determines  the  sign  of  \(\cn(z,k')\).  The
algorithm is deemed to have converged when \(b\) ceases to increase: this works
whatever floating point precision \texttt{FLOAT} is specified.}
\label{fig:arithgeom}
\end{figure}

\subsection{Evaluation of {\zolotarev} Coefficients} \label{sec:evaluate}

The arithmetico-geometric  mean lets  us evaluate {\jacobi}  elliptic functions
for real arguments  \(z\) and real parameters \(0<k<1\).  For complex arguments
we  can  use the  addition  formula to  evaluate  \(\sn(x+iy,k)\)  in terms  of
\(\sn(x,k)\) and \(\sn(iy,k)\), and the  latter case with an imaginary argument
may  be  rewritten in  terms  of  real  arguments using  {\jacobi}'s  imaginary
transformation. We  can either use  these transformations to  evaluate elliptic
functions of  complex argument numerically,  or to transform  algebraically the
quantities we wish to evaluate into  explicitly real form. Here we shall follow
the latter approach, as it is  more efficient to apply the transformations once
analytically.

\zolotarev's formula (\ref{eq:E}) is \[R(x) = {2\over1+{1\over\lambda}} {x\over
kM}  \prod_{m=1}^{\lfloor\half n\rfloor}  {k^2 -  c_mx^2\over k^2  - c'_mx^2}\]
with \(c_m \defn \sn(2iK'm/n,  k)^{-2}\) and \(c'_m \defn \sn(2iK'(m-\half)/ n,
k)^{-2}\).  We may  evaluate the  coefficients  \(c_m\) and  \(c'_m\) by  using
{\jacobi}'s  imaginary transformation~(\ref{eq:F}),  \[ c_m  = -\left[{\cn(2K'm
/n},k')   \over   \sn(2K'm/n,k')\right]^2,   \quad   c'_m   =   -\left[{\cn(2K'
(m-\half)/n,k') \over \sn(2K'(m-\half)/n,k')}\right]^2.\] We also know that \(M
= \prod_{m=1}  ^{\lfloor\half n\rfloor} (1-c_m)/(1-c'_m)\), and  \( 1/\lambda =
(\bar\xi/M)  \prod_{m=1}^{\lfloor\half n\rfloor}  (1-c_m  \bar\xi^2) /  (1-c'_m
\bar \xi^2)\)  with \(\bar\xi \defn  \sn(K+iK'/n,k)\). We may use  the addition
formula  to express  the {\jacobi}  elliptic functions  of complex  argument in
terms of ones with purely real or imaginary arguments, so
\begin{eqnarray*}
\bar\xi &=& \sn\left(K + {iK'\over n}, k\right) = {\sn K \cn{iK'\over n}
\dn{iK'\over n} + \sn{iK'\over n} \cn K \dn K\over1 - \left(k \sn K
\sn{iK'\over n}\right)^2}\\ &=& {\cn{iK'\over n} \dn{iK'\over n}\over1 -
\left(k\sn{iK'\over n}\right)^2} = {\cn{iK'\over n}\over \dn{iK'\over n}}.
\end{eqnarray*}
These may be converted to expressions involving only real arguments by the use
of \jacobi's imaginary transformation~(\ref{eq:F}),
\begin{eqnarray*}
\sn\left({iK'\over n}, k\right) &=& {i\sn\left({K'\over n},k'\right) \over
\cn\left({K'\over n},k'\right)},\\ \cn\left({iK'\over n},k\right) &=& \sqrt{1 +
\left[{\sn\left({K'\over n},k' \right)\over \cn\left({K'\over
n},k'\right)}\right]^2} = {1\over \cn\left({K' \over n},k'\right)},\\
\dn\left({iK'\over n},k\right) &=& \sqrt{1 + \left[{k \sn\left({K'\over
n},k'\right)\over \cn\left({K'\over n},k'\right)}\right]^2} =
{\dn\left({K'\over n},k'\right) \over\cn\left({K'\over n},k'\right)},
\end{eqnarray*}
giving the simple result \(\bar\xi = 1/\dn(K'/n,k')\).

Putting these results together we have \[R(x) = Ax\prod_{m=1}^{\lfloor\half
n\rfloor} {x^2-a_m\over x^2-a'_m}\] with
\begin{eqnarray*}
a_m  =  {k^2\over  c_m}  =  -  \left[k  {\sn\left({2K'm\over  n},k'\right)\over
\cn\left({2K'm\over  n},k'  \right)}\right]^2,&& a'_m  =  {k^2\over  c'_m} =  -
\left[k  {\sn\left({2K'(m  -  \half)\over n},k'\right)\over  \cn\left({2K'(m  -
\half)\over  n},k'\right)}  \right]^2,\\   A  =  {2\over1+1/\lambda}{1\over  k}
\prod_{m=1}^{\lfloor\half   n\rfloor}  {c_m\over  c'_m}\left({1-c'_m\over1-c_m}
\right),&& \Delta = {1-\lambda\over1+\lambda},
\end{eqnarray*}
where \(\Delta\) is the maximum error of the approximation.

\section*{Acknowledgements}

I would like to thank Urs Wenger for helpful discussions, and Wasseem Kamleh
for correcting the quite unnecessarily obfuscated and previously incorrect bit
manipulations in Fig.~\ref{fig:sncndnK}.

\end{document}